\RequirePackage{ifpdf}
\documentclass{PoS}

\usepackage{graphics, graphicx, url}
\usepackage{natbib}
\bibpunct{(}{)}{;}{a}{}{,}

\title{Very Long Baseline Interferometry with the SKA}

\ShortTitle{VLBI with the SKA}

 \author{\speaker{Zsolt Paragi}$^{1}$, 
      Leith Godfrey$^{2}$, 
      Cormac Reynolds$^{3}$,
      Maria Rioja$^{4,5}$,
      Adam Deller$^{2}$,
      Bo Zhang$^{6}$,  
      Leonid Gurvits$^{1,37}$,
      Michael Bietenholz$^{7,38}$,
      Arpad Szomoru$^{1}$,
      Hayley Bignall$^{3}$,
      Paul Boven$^{1}$,
      Patrick Charlot$^{8}$, 
      Richard Dodson$^{9,10}$, 
      S\'andor Frey$^{11}$, 
      Michael Garrett$^{2,12}$, 
      Hiroshi Imai$^{13}$, 
      Andrei Lobanov$^{14}$,
      Mark Reid$^{15}$, 
      Eduardo Ros$^{14,39}$, 
      Huib van Langevelde$^{1,12}$, 
      J. Anton Zensus$^{14}$, 
      Xing Wu Zheng$^{16}$,
      Antxon Alberdi$^{17}$,
      Iv\'an Agudo$^{1}$,
      Tao An$^{6}$,
      Megan Argo$^{18}$,
      Rob Beswick$^{18}$,
      Andy D. Biggs$^{19}$,
      Andreas Brunthaler$^{14}$,
      Robert M. Campbell$^{1}$,
      Giuseppe Cim\'o$^{1}$,
      Francisco Colomer$^{4}$,
      St\'ephane Corbel$^{20}$,
      John Conway$^{21}$,              
      David Cseh$^{22}$,
      Roger Deane$^{23,33}$,
      Heino Falcke$^{22}$,
      Krisztina Gab\'anyi$^{41,11}$,
      Marcin Gawronski$^{24}$,              
      Michael Gaylard$^{7}\footnote{Deceased}$,
      Gabriele Giovannini$^{25}$,
      Marcello Giroletti$^{25}$,
      Ciriaco Goddi$^{1}$,
      Sharmila Goedhart$^{7}$,
      Jos\'e L. G\'omez$^{17}$,
      Alastair Gunn$^{18}$,
      Taehyun Jung$^{9}$,
      Preeti Kharb$^{26}$,
      Hans-Rainer Kl\"ockner$^{14}$,
      Elmar K\"ording$^{22}$,
      Yurii Yu. Kovalev$^{27,14}$,
      Magdalena Kunert-Bajraszewska$^{24}$,   
      Michael Lindqvist$^{21}$,
      Matt Lister$^{28}$,
      Franco Mantovani$^{14,25}$,
      Iv\'an Mart\'i-Vidal$^{21}$,
      Mar Mezcua$^{29}$,
      John McKean$^{2}$,
      Enno Middelberg$^{30}$,
      James Miller-Jones$^{3}$,
      Javier Moldon$^{2}$,
      Tom Muxlow$^{18}$,
      Tim O'Brien$^{18}$,
      Miguel P\'erez-Torres$^{17}$,
      Sergei Pogrebenko$^{1}$,
      Jonathan Quick$^{7}$,
      Anthony P. Rushton$^{31,40}$,
      Richard Schilizzi$^{18}$,
      Oleg Smirnov$^{32,33}$,
      Bong Won Sohn$^{9}$,
      Gabriele Surcis$^{1}$,
      Greg Taylor$^{34}$,
      Steven Tingay$^{3}$,
      Valeriu Tudose$^{35}$,
      Alexander van der Horst$^{36}$,
      Joeri van Leeuwen$^{2,36}$,
      Tiziana Venturi$^{25}$,
      Ren\'e Vermeulen$^{2}$,
      Wouter Vlemmings$^{21}$, 
      Aletha de Witt$^{7}$,        
      Olaf Wucknitz$^{14}$,
      Jun Yang$^{21}$  
      \\
      $^{1}$JIVE, $^{2}$Astron, $^{3}$ICRAR-Curtin~U., $^{4}$OAN, $^{5}$ICRAR-UWA, 
      $^{6}$ShAO, $^{7}$HartRAO, $^{8}$U.~Bordeaux, $^{9}$KASI, $^{10}$ICRAR-UWA, 
      $^{11}$F\"OMI~SGO, $^{12}$ U. Leiden, $^{13}$Kagoshima~U., $^{14}$MPIfR, 
      $^{15}$Harvard-Smithsonian CfA,  $^{16}$SASS-Nanjing~U., $^{17}$IAA-CSIC, 
      $^{18}$JBO-U. Manchester, $^{19}$ESO, $^{20}$CEA Saclay, $^{21}$OSO, $^{22}$RU~Nijmegen, 
      $^{23}$U. Cape Town, $^{24}$Toru\'n CfA--Nicolaus Copernicus U., $^{25}$IRA--INAF, 
      $^{26}$IIA Bangalore, $^{27}$ASC--Lebedev Phys. I., $^{28}$Purdue U., $^{29}$IAC, 
      $^{30}$Ruhr-U. Bochum, $^{31}$U. Oxford, $^{32}$ Rhodes U., $^{33}$ SKA South Africa, 
      $^{34}$UNM, $^{35}$Inst. Sp. Sci., $^{36}$U. Amsterdam, $^{37}$ TU Delft, 
      $^{38}$ York U., Toronto, $^{39}$ U. Valencia, $^{40}$ U. Southampton, $^{41}$ U. Szeged
      \\
      E-mail: \email{zparagi@jive.nl}
      }

\abstract{Adding VLBI capability to the SKA arrays will greatly broaden 
the science of the SKA, and is feasible within the current specifications. 
SKA-VLBI can be initially implemented by providing phased-array outputs for 
SKA1-MID and SKA1-SUR and using these extremely sensitive stations with other
radio telescopes, and in SKA2 by realising a distributed configuration 
providing baselines up to thousands of km, merging it with existing 
VLBI networks. The motivation for and the possible realization of SKA-VLBI is 
described in this paper.} 

\FullConference{
Advancing Astrophysics with the Square Kilometre Array\\
June 8-13, 2014\\
Giardini Naxos, Italy}

\begin{document}

\section{Introduction}

A high angular resolution capability has long been considered an essential part of 
the Square Kilometre Array (SKA) concept 
\citep{garrett00, gurvits04, fomreid04, schilizzi07, godfrey12}.
Very long baseline interferometry with the SKA (hereafter SKA-VLBI) will provide
very sensitive, milliarcsecond (mas) resolution imaging that is important, for example, 
to study active galactic nuclei (AGN) down to very low luminosities,  
to understand the detailed physics of jet formation and its coupling 
to the accretion process, as well as the growth of the first generation of massive
black holes in the universe \citep{Agudo14}, and their role in regulating
star formation \citep[a.k.a. feedback processes, see][]{Prandoni14,Morganti14}.
The detection of a great number of tidal disruption events (TDE), and radio imaging 
at mas-scale, will be invaluable for the understanding of jet formation in a pristine 
environment, and will possibly reveal a new population of massive black holes 
(MBH, $M_{\rm BH}~\sim10^4-10^6 M_{\odot}$)
that may resemble the black hole seeds in the early universe \citep{Donna14}.
High-fidelity SKA-VLBI polarimetric observations will constrain the magnetic field
structure close to the jet launch site, providing important constraints on the jet 
launch mechanism and related processes \citep{BZ77,BP82}. In connection with
the Cherenkov Telescope Array (CTA), it will help reveal the nature of the large 
population of hitherto unidentified high-energy sources \citep{Giroletti14}.
The various \lq\lq exotic'' accreting black hole systems that will potentially be
revealed by mas-scale deep SKA-VLBI imaging include low-power MBH in the centres of dwarf 
galaxies \citep{Paragi14}, off-centre intermediate-mass black holes 
\citep[$M_{\rm BH}~\sim10^2-10^4 M_{\odot}$,][]{Wolter14}, 
and dual- or multiple supermassive black holes\footnote{For two recently reported 
candidates awaiting confirmation see e.g. \citet{Gitti13,Deane14b}} 
\citep[SMBH, $M_{\rm BH}~\sim10^6-10^{10} M_{\odot}$,][]{Deane14a}
in advanced stages of mergers.  
It will allow the detection of $10^6$~$M_{\odot}$ dark matter haloes from high resolution 
imaging of gravitationally lensed arcs to investigate galaxy formation scenarios. It will 
also enable a test of models for dark energy from the measurement of geometric distances 
to high-redshift galaxies with nuclear water masers \citep{McKean14}. 

Ultra-precise astrometry at the microarcsecond level 
to determine distances and transverse velocities via the measurement of proper motions 
and parallaxes of Galactic objects will be possible out to a distance of tens of kpc. 
Achieving this for a large fraction of the radio pulsar population detected in the 
SKA Galactic pulsar census will enable  
strong field tests of gravity in a broad range of relativistic binary systems 
\citep{Kramer14,Shao14}, the detection of the gravitational wave background
\citep{Janssen14}, tomographic modelling of the large scale Galactic magnetic field 
and mapping the ionized interstellar plasma in the Galaxy \citep{Han14},
and constraining the physics of neutron stars \citep{Watts14}. 
Curiously, the gravitational wave background might be constrained independently by observing
its subtle effect on the apparent position of quasars 
\cite[e.g.][see also Sect.~\ref{sec:abs_astr}]{Jaffe04}.
Ultra-precise astrometry will be very important 
for many other classes of compact radio sources with emission in the SKA 
frequency range -- for instance, masers \citep{Green14}, protostellar objects 
\citep[e.g.][]{loinard07a},  
and a variety of accreting stellar objects such as novae, isolated black holes, 
or accreting neutron stars and black holes in binary systems \citep[a.k.a. microquasars,][]{Corbel14}.
 Proper motion measurements of stellar-mass black holes will provide constraints 
 on models of their formation, while measuring their parallax distances 
 will provide accurate luminosities which are important for tests of accretion 
 physics \citep[e.g.][]{miller-jones09a}. 
For radio-emitting stars, in addition to the distance via parallax, the presence of 
a planetary companion can be sought via the reflex motion of the star \citep[e.g.][]{guirado07a, bower09a}.
The excellent sensitivity on very long baselines will mean that extragalactic stellar
explosions like supernovae \citep{Bietenholz2008,Perez-Torres14} and gamma-ray burst 
afterglows \citep[GRB,][but see also Sect.~\ref{sec:GRBjets}]{Burlon14} will be imaged, in total 
intensity as well as in polarization, in more detail than is possible today.

\begin{figure} 
\centering
\includegraphics[scale=0.45]{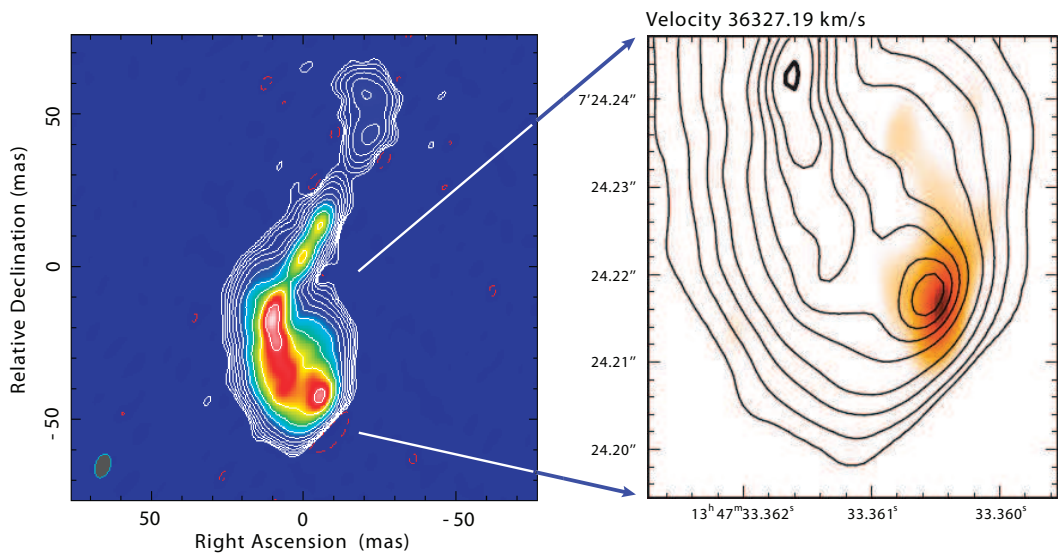}
\caption{Left: global VLBI continuum image of the young, restarted AGN 4C12.50 at 1.27~GHz.
Right: H{\sc I} absorption (orange-white) superimposed on the continuum image (contours)
of the southern lobe. The deepest H{\sc I} absorption is co-spatial with the termination 
point of the jet; there is extended H{\sc I} absorption detected as well. The jet drives 
a large-scale atomic (H{\sc I}) and cold molecular (CO) outflow. This result demonstrates 
that jet-driven outflows may play an important role in AGN feedback mechanisms 
\citep{morganti13}. 
\label{fig:4C12.50}}
\end{figure}

A number of high impact scientific
results produced in the last few years demonstrate the potential of very 
sensitive and flexible SKA-VLBI with mas-scale imaging capability. These include  
global VLBI H{\sc I} spectral line imaging of the AGN 4C12.50 (see Fig~\ref{fig:4C12.50}), 
demonstrating how AGN jets drive large-scale outflows and thus contribute to AGN 
feedback \citep{morganti13}; an accurate parallax measurement of SS~Cyg in a series of 
triggered VLBI experiments (see Fig~\ref{fig:parallax}), proving that the 
disc instability model for accretion is correct \citep{miller-jones13};  
and the recent finding that binary orbits may be the driver of $\gamma$-ray emission and 
mass ejection in classical novae \citep{Chomiuk14}.
To fully explore the broad range of science cases we refer the reader to the various
chapters mentioned above, as well as \citet{godfrey11, godfrey12}. In this paper we focus 
on practical SKA-VLBI issues and highlight only a few SKA-VLBI science applications.

We describe possible realisations of SKA-VLBI in Sect.~\ref{sec:configs}, along
with expected sensitivities based on the SKA1 Baseline Design. Sect.~\ref{sec:calibration}
describes calibration requirements and explains the need for  
forming multiple phased-array beams (in one or more sub-arrays) and why 
simultaneous production of
phased-array and local interferometer data is necessary. Astrometry with SKA-VLBI
will be described in Sect.~\ref{sec:astrometry}. In Sect.~\ref{sec:GAIA} we will show how
{\it Gaia} 
astrometry can be improved using SKA-VLBI, and how these complementary facilities
may open up new fields of research. This will be followed by introducing the need and
potential for SKA-VLBI surveys in Sect.~\ref{sec:surveys}. Highlighting the superior
resolution of SKA-VLBI even for very faint targets, we will show in Sect.~\ref{sec:GRBjets}
that --in certain cases-- it will be possible to resolve extreme GRB afterglows while 
these are still in the ultra-relativistic regime. Various issues with correlation and 
data transport will be described in Sect.~\ref{sec:data}.

\section{VLBI configurations of SKA} \label{sec:configs} 

There is a significant difference between the planned operation of the high
angular resolution component for SKA1 and SKA2: in SKA1, the SKA core will
be operated as a sensitive element (or elements) to be added to existing
VLBI networks, increasing the sensitivity but not the resolution of those
networks.  In this case, the \lq\lq remote stations'' comprise existing
facilities, and the SKA will participate as the most sensitive element in
an otherwise conventional VLBI array (see Table~\ref{tab:ska-vlbi}). Key to
the success of this approach will be the addition of a few (2--4) remote
stations in Africa. Ideal locations for these stations would be the
developing African VLBI Network (AVN) stations in Zambia, Ghana, Kenya and
Madagascar \citep{Gaylard+2011}.  These additional stations will provide
the short and medium length baselines to the SKA core to give good
$uv$-coverage.

\begin{figure} 
\centering
\includegraphics[scale=0.40, angle=-90,
clip]{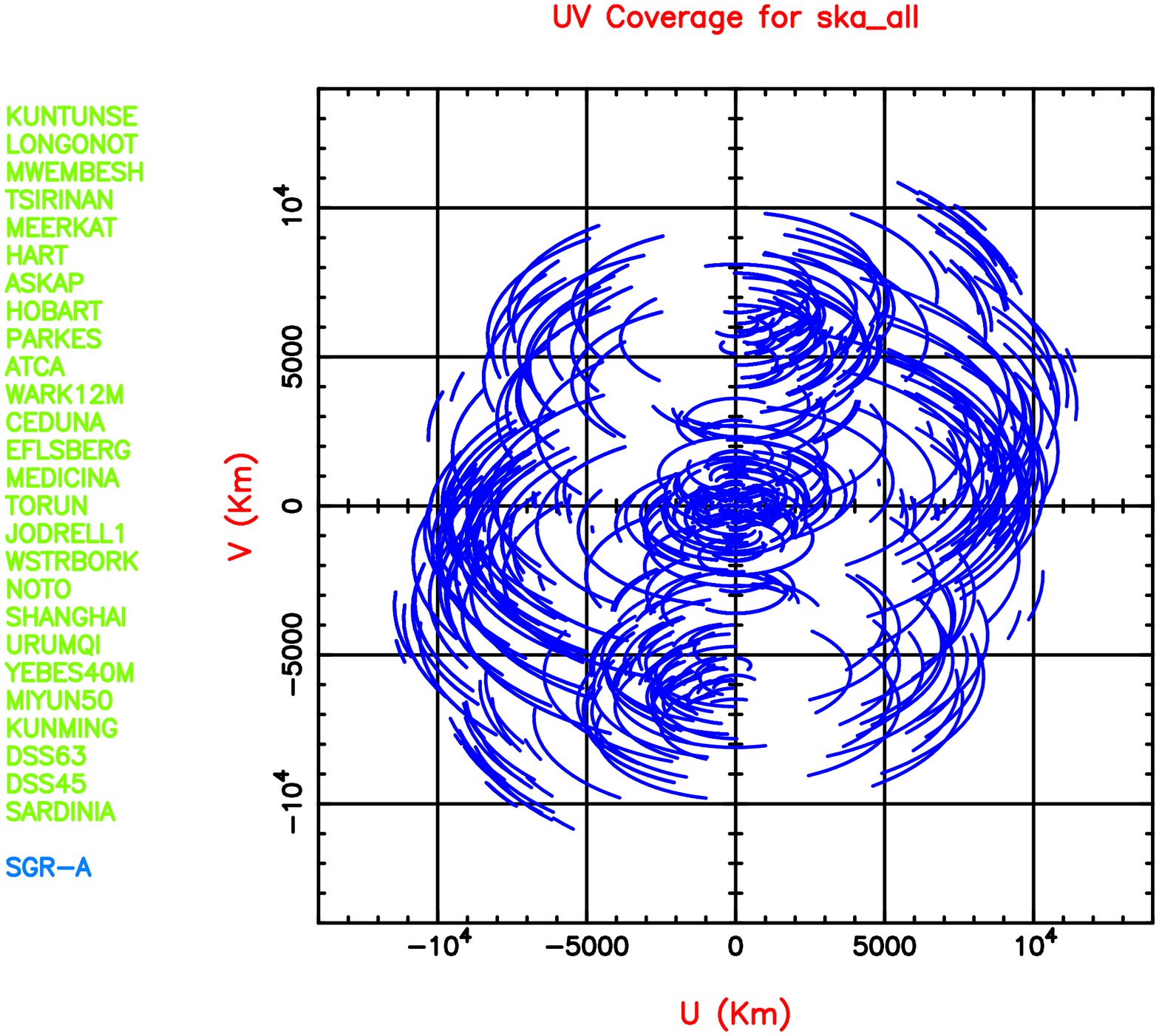}
\includegraphics[scale=0.40, angle=-90,
clip]{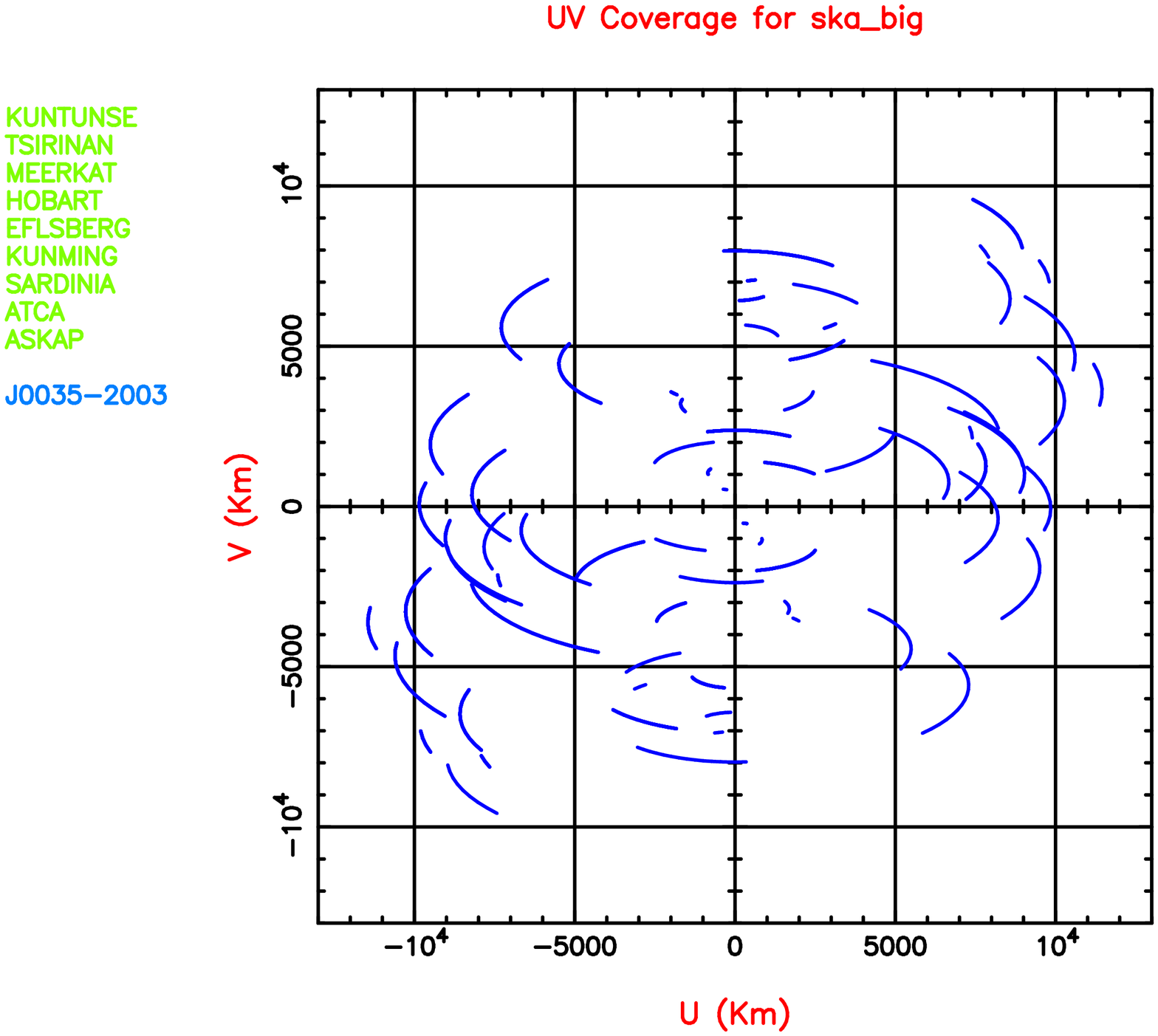}
\caption{Left: 
$uv$-coverage of the Galactic Centre in a 12-hours observation with 
24 telescopes combined in the EVN, CVN, LBA and the AVN with SKA1-MID. 
Right: 
$uv$-coverage of a more typical 4-hours observation of a DEC $-20^{\circ}$ 
source with 9 telescopes selected from the above configuration. 
\label{fig:uv}}
\end{figure}

The left panel of Figure~\ref{fig:uv} shows the single-frequency, 12-hours
$uv$-coverage that could be obtained at the Galactic centre with a global
imaging array including a large number of telescopes. The right panel shows
a more typical $uv$-coverage for a 4-hours track on a source at declination
$-20^{\circ}$ using just a handful of the more sensitive telescopes
available in the global array, but still providing baseline lengths up to
10,000~km in length. In the coming era of rapidly expanding e-VLBI
capabilities the realisation of such an array should not present any
significant logistical problems. Coordinated proposals and observations are
already offered by the European VLBI Network (EVN), the Chinese VLBI
Network (CVN), the Japanese VLBI Network (JVN), the Korean VLBI Network
(KVN), the Long Baseline Array (LBA) the Very Long Baseline Array (VLBA),
and the High Sensitivity Array (HSA).  Table~\ref{tab:ska-vlbi} gives the
likely sensitivity of arrays that could be formed with SKA1 era
assuming various configurations and bandwidths.

\begin{table}
\begin{tabular}{lrrrrr}
\noalign{\smallskip}
\hline
\hline
\noalign{\smallskip}
SKA Band & SKA-core & Bandwidth & Remote tel.& Baseline sens.& Image noise   \\
         & SEFD [Jy]& [MHz]     & SEFD [Jy]  & 60s [$\mu$Jy] & 1hr [$\mu$Jy/beam] \\
\noalign{\smallskip}
\hline
\hline
\noalign{\smallskip}
50\% SKA1-MID & 5.2 & 256       & 20         & 82            & 9 \\  
SKA1-MID      & 2.6 & 1024      & 20         & 29            & 3 \\  
Full SKA      & 0.26& 2048      & 20         & 3             & 0.05 \\  
\hline
\noalign{\smallskip}
\end{tabular}
\caption{Typical expected 1$\sigma$ baseline and image sensitivities of various SKA-VLBI 
configurations at $\sim$3--8 GHz, with the inner 4 km of SKA core phased up. All the 
baseline sensitivities are given for a 100m-class remote telescope. 50\% SKA1-MID (early 
operations): assuming an accompanying array of 5 25--30m dishes and a 100m-class antenna. 
SKA1-MID -- same configuration. Note at $\sim$1--3 GHz and including SKA1-SUR as well will 
provide a similar sensitivity. Full SKA: 10x more sensitive than SKA1-MID. 
\label{tab:ska-vlbi}
}
\end{table}

There are, however, limitations to the approach described above  \citep{garrett00}. 
To address these limitations, in SKA2 the remote stations will be an 
integrated part of the array, with up to 25\% of the collecting area distributed 
along spiral arms in remote stations, extending thousands of kilometres from the 
core, and the SKA will effectively operate as a real-time electronic-VLBI (e-VLBI) array. 
Merging with telescopes from existing VLBI networks will still be possible; note that in 
this case the SKA processor will have to act as the VLBI correlator (Table~\ref{tab:ska-vlbi}).
Many of the technical requirements for implementation of SKA-VLBI in phase~1 will 
also be requirements for phase~2, and as such, implementation of the high angular
resolution capability in phase~1 will be an essential step to realising the
high angular resolution component in phase~2.

Even in the early science phase (50\% SKA1-MID sensitivity) SKA-VLBI will match or 
exceed the current capabilities of the typical e-EVN 
configuration\footnote{The e-EVN consists of EVN telescopes with real-time e-VLBI capability,
streaming data at a rate of 1~Gbit~s$^{-1}$. Here we assumed the e-EVN without Arecibo.
See \url{http://www.evlbi.org/evlbi/e-vlbi_status.html}}, but will be 
capable of accessing the entire southern sky, including tracking the Galactic center for 
many hours! SKA-VLBI with the full capacity SKA1-MID and/or SKA1-SUR, especially when adding 
FAST \citep{Nan11}, will outperform current global VLBI arrays including the most sensitive 
telescopes.
While we are not considering it further here, we note that VLBI with SKA1-LOW 
($\nu$<350 MHz) has the potential of using speckles of interstellar scattering 
to achieve sub-nano-arcsec resolution in certain cases. With current instruments 
this is possible for some bright pulsars \citep{Pen14}, but the sensitivity of 
the SKA is required for more general applications.

\section{SKA-VLBI calibration} \label{sec:calibration}

Calibration of the VLBI datastream from the phased-up core of the SKA
array would follow much the same methods as currently used for phased 
arrays in VLBI. When connected element interferometers such as the 
Australia Telescope Compact Array (ATCA), The Karl G. Jansky Very Large Array (VLA) 
or the Westerbork Synthesis Radio Telescope (WSRT) are operated as an element 
of a VLBI network, the internal calibration of the array is used to measure the 
gains for the individual array antennas. 
The signals from these are then scaled and summed in a tied array 
output from the correlator. We assume that the SKA will function along these
lines as part of VLBI operations. Besides forming a tied array beam 
(or rather beams, see below), another crucial requirement is the provision of 
the metadata, such as the system temperature 
and weather information, by the array. The gains of the phased-array sum 
need to be accurately determined and recorded. The metadata will be essential 
for the accurate conversion of the VLBI correlation into a measure of 
correlated flux density. These calibrations are particularly important as 
such data will be dominated by the baselines to the phased SKA because of 
the large weights these baselines will have due to the large collecting area.

One issue in VLBI is that primary flux density calibrators are not 
available to further improve on a-priori amplitude calibration.
Primary calibrators are resolved on mas scales, while sources that
are compact on baselines up to $\sim$10,000~km are variable. SKA-VLBI
will offer a great solution to this by providing local interferometer
and phased-array data simultaneously. The flux densities and
polarization properties of compact calibrators can be measured
using the local interferometer data during the VLBI observations, 
leading to very accurate flux density and polarization
calibration (for both polarization leakage and polarization position
angle) of the VLBI data product.

\subsection{Multi-view calibration}

To provide 10~$\mu$as astrometric accuracy using single source phase referencing at 
1.6 GHz one would require an extremely nearby reference source. One can estimate from 
the expressions in \citet{a07} that the reference sources should be no more than 
60 arcsec from the target to provide the required accuracy. The projected source 
counts suggest that there will not be sufficient calibrator source density to 
provide this, even at the sensitivities of SKA1-MID to SKA1-SUR baselines 
(see Table\ref{tab:ska-vlbi}). 

However, using lines of sight to multiple calibrator sources (minimally 3) it is possible to 
solve for a full 2D correction to the spatial atmospheric distortions around the VLBI target, 
which will provide significantly improved calibration, imaging and 
astrometric precision compared to that from a single calibrator
\citep[see e.g.][and references therein]{rioja09}.
This is particularly important in the ionosphere-dominated regime ($<$5~GHz).
The improvement arises from the fitting of a spatial function to the
calibration residuals and interpolating this model to the target
position, which allows a more accurate calibration solution in the 
target direction.  
Simulations show \citep{sergio} that by using Multi-view approaches 
one can achieve an order of magnitude better astrometric accuracy 
than by using a single calibrator. The necessary calibrator-target separation 
is $\lesssim$5 arcminutes, and the probability of detecting suitable calibrators 
at such distances seems promising at least up to 5~GHz \citep{godfrey11}, making 
this band the best for SKA-VLBI astrometry (described below).

When SKA1-MID and SKA1-SUR are phased for use in SKA-VLBI, the 
resultant tied-array beams will be narrow ($\sim$10 arcseconds
at 1.6 GHz if the inner 4 km of the core is phased up). This is much narrower than
the field-of-view of the 25-100m class telescopes which will form the remainder
of the SKA1-VLBI array, and much smaller than the separation between
potential SKA-VLBI calibrator sources.  Accordingly, a different
SKA tied array beam will typically be needed per calibrator source.
This drives the requirement for a minimum of 4 SKA1-MID and SKA1-SUR
VLBI beams.

\section{Astrometry} \label{sec:astrometry}

\subsection{Differential astrometry}

\begin{figure} 
\centering
\includegraphics[scale=0.33, viewport=18 0 594 412, angle=0,
clip]{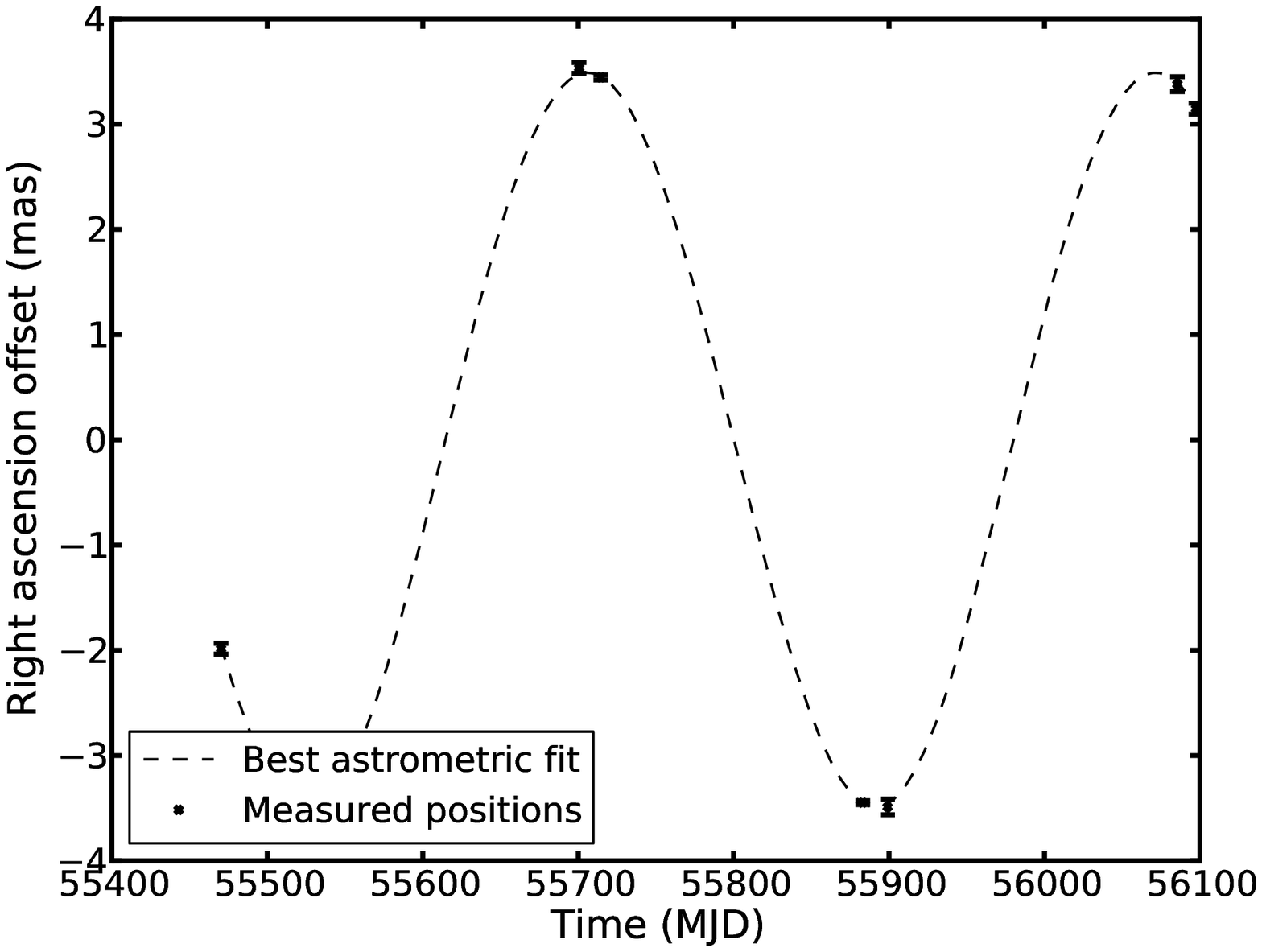}
\includegraphics[scale=0.42, viewport=0 0 503 311, angle=0, clip]{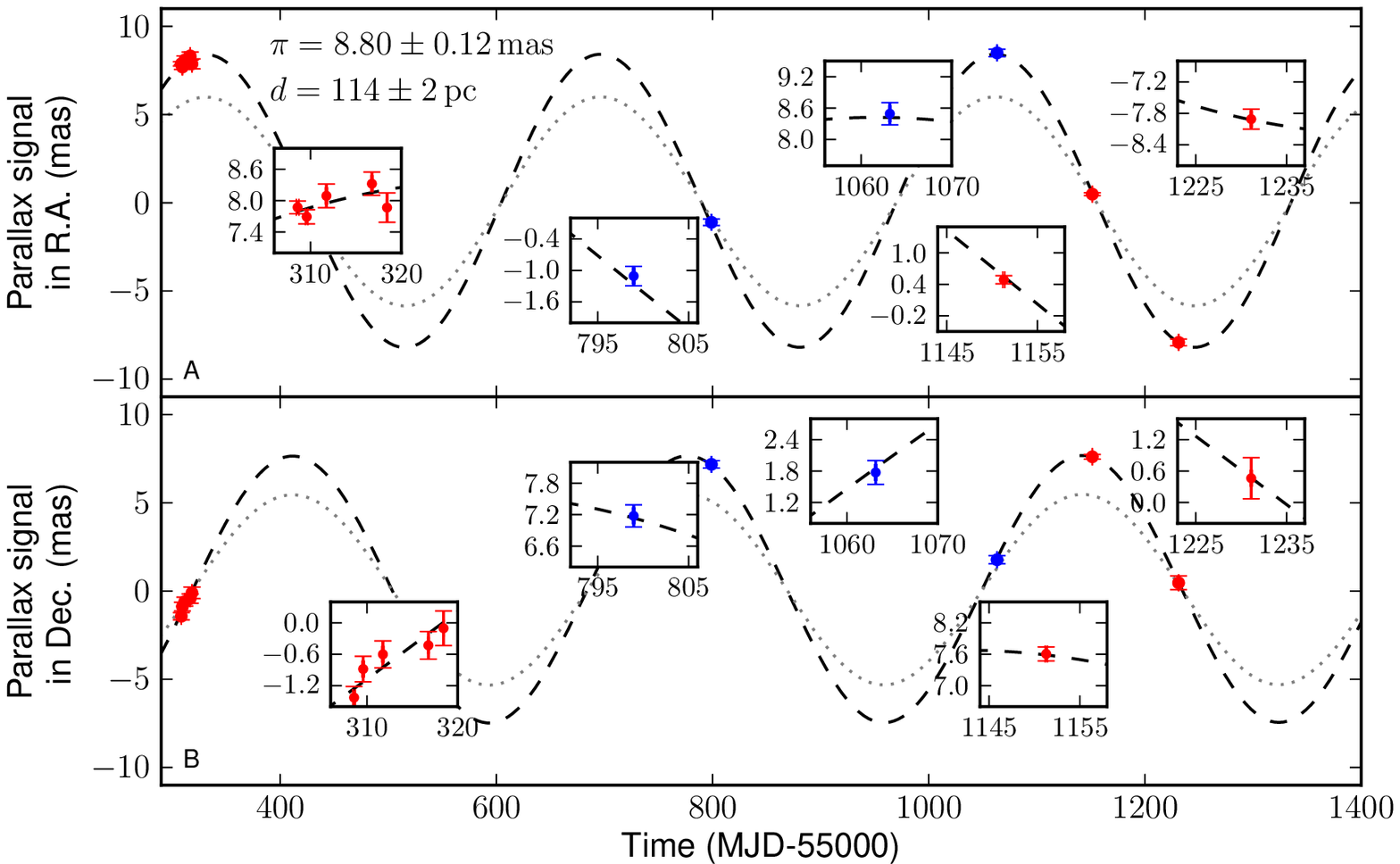}
\caption{Left: the proper motion and parallax of the binary pulsar J2222-0137 as 
measured by the VLBA. The distance to this source was measured to 0.4$\%$ 
accuracy to be 267.3 pc, using 7 observations spread over 20 months
\citep[][Fig.~5]{deller13a} 
Right: VLBA and e-EVN accurate parallax distance measurement 
of the dwarf nova SS~Cyg from a series of triggered VLBI experiments,
which removed a major challenge to our understanding of accretion theory 
\citep[][Fig.~2]{miller-jones13}
 \label{fig:parallax}}
\end{figure}

Differential VLBI astrometry is obtained by carefully calibrating standard phase-referenced 
VLBI observations \citep[see][and references therein]{reid14}. The calibration is derived 
from one source (or more sources, using Multi-view calibration) registered within the 
International Celestial Reference Frame 
(ICRF), and the same local frame is constructed over multiple observations, meaning that 
systematic contributions to positional errors remain constant to first order. Thus, while 
the absolute position of the target will be in error by a small but unknown amount, the 
changes between epochs -- including, importantly, annual geometric parallax and source 
proper motion -- are reliable.  Compared to absolute astrometry (discussed below), 
differential astrometry can obtain higher relative precision and can target much fainter 
objects. Differential astrometry with VLBI provides the highest precision direct distance 
measurements of objects outside the solar system available in astronomy (see Figure 
\ref{fig:parallax}), and as such it contributes a crucial rung to the distance ladder.

The attainable accuracy with differential astrometry can be limited by 4 factors: 
1) the noise-limited position fit of the target (determined by array resolution and 
sensitivity and target brightness); 2) the registration of the target within the 
calibrator frame (limited by the proximity of the calibrator(s) and the calibration 
solution interval); 3) the stability of the calibrator frame itself (limited by the 
intrinsic nature of the source(s) used); and, in case of parallax measurements 
4) the stability of the target emission centroid. The participation of SKA1-MID and 
SKA1-SUR in VLBI astrometric observations will lead to considerable reductions in 
the sum of these error contributions; so much so that the expected noise floor is 
difficult to extrapolate from present observations.

The instantaneous sensitivity of a VLBI array containing SKA1-MID and SKA1-SUR would 
be more than a factor of 10 better than the VLBA. 
This translates into a corresponding reduction in the first error component. 
Moreover, it means that 
fainter sources can be used as calibrators. As explained in Sect.~\ref{sec:calibration},
several calibrator sources within $\sim$5 arcminutes could be expected, that will result in
up to an order of magnitude reduction in the second error component (which is usually 
the limiting factor with current astrometric observations). By virtue of the fact that 
many sources will be used to construct the calibrator frame, cross-checks will be 
possible allowing the removal of sources which demonstrate discernible structure 
evolution, mitigating the third error component. This is currently poorly constrained 
\citep{Fomalont11}, as it is below the error floor for most current observations. 
The fourth error component is only relevant for certain targets (such as masers) and 
can be mitigated by compressing the campaign duration.

Astrometric observations with current instruments are capable of reaching parallax precisions
of $\sim$10~$\mu$as \citep[e.g.][]{deller13a, nagayama11a,Zhang13,Reid11}. SKA-VLBI
has the potential to reach parallax accuracies of 3~$\mu$as or better,
sufficient for a precise distance to any Galactic object along a line of sight that is not
substantially affected by scattering.

\subsection{Absolute astrometry and geodesy} \label{sec:abs_astr}

Improving the reference frames - both celestial and terrestrial - that are needed for 
differential astrometry is the domain of absolute astrometry and geodesy. Absolute 
astrometry can also be used to study very small secular effects such as the 
\lq\lq Galactic aberration'' introduced by the acceleration of the 
Solar System barycentre in the Galactic potential, 
and apparent quasar proper motions introduced by the low-frequency gravitational wave
background \citep[e.g.][and references therein]{gwinn97,titov11}.
For these observations, one cannot make use of a 
nearby calibrator or calibrators to largely eliminate the effect of the ionosphere or 
troposphere, and so they dominate the error and must be modeled as accurately as possible. 
As long as the spanned bandwidth is relatively wide, the ionospheric contribution can be 
precisely measured and compensated because it introduces a dispersive delay. To model the 
troposphere, observations of many sources at different locations are needed within a short
period of time -- ideally minutes. The SKA can provide multiple subarrays, but the efficiency
will be dictated by the number of non-SKA antennas available to participate.

Unless the number of VLBI-capable antennas in the southern hemisphere is considerably 
expanded, the addition of SKA1-MID will not result in large improvements in the accuracy of 
the ICRF. However, SKA2 with intrinsic baselines up to thousands of kilometres as 
well as higher frequencies, will be able to make a significant improvement over the current 
ICRF accuracy. To achieve that, it will be necessary to correct for the opacity-shift in 
AGN jets for the ICRF defining sources \citep[e.g.][and references therein]{Paragi00,Porcas12}, 
which will also be crucial for aligning the radio and the {\it Gaia} optical reference frames 
as explained below.

\section{Synergies with Gaia} \label{sec:GAIA}

\subsection{Calibrating Gaia parallaxes}

{\it Gaia} --- the successor to Hipparcos --- was launched in the end of 2013,
with an ambitious aim to chart a three-dimensional map of the Milky Way.
{\it Gaia} will use twin telescopes to observe two regions in the
sky simultaneously to reach the expected parallax accuracy of $\sim$20~$\mu$as 
at magnitude $\sim$15. However, even very small periodic variations in the
so-called basic angle between the two fields of view could lead to an
undesirable global offset of the measured
parallaxes~\citep{Mignard11}.

Although the variations are monitored by an on-board metrology system, it is 
important to verify the {\it Gaia} parallaxes by independent methods. 
Among methods which are in principle available to verify the {\it Gaia} parallaxes, 
distant quasars with practically zero parallaxes are the most promising candidates, 
but the parallax zero
point determined with this method would introduce a possible additional bias
from foreground stars contaminating the sample~\citep{Windmark11}.
Parallax measurements of stars with accuracy comparable to that of {\it Gaia} are
possible with VLBI. However, due to the limited sensitivity, only a few
stars with optical counterparts observable with {\it Gaia} have been
measured with VLBI~\citep[e.g.][]{miller-jones13,Ratner12}. The
number of objects detectable with current VLBI arrays is not high enough
to calibrate {\it Gaia}, therefore we will need SKA-VLBI to accurately
measure parallaxes for a significant number of Galactic {\it Gaia} targets.

As can be seen in Table~\ref{tab:ska-vlbi}, the image sensitivity
of a moderate SKA-VLBI array is 3~$\mu$Jy beam$^{-1}$~hr$^{-0.5}$ at $\sim3-8$~GHz.
Assuming a baseline length up to 10,000 km (resolution $\sim$1~mas), this will 
ensure a noise-limited position fit with a theoretical precision of a few $\mu$as 
for stars with flux density exceeding $\sim$1~mJy (SNR>100), even if observed in 
a snapshot mode ($\sim$10~min. per target). 
As explained in previous sections, the systematic errors can be
reduced significantly by using multiple calibrators within a few arcminutes. 
For verification of {\it Gaia} parallaxes, we need to observe radio targets
which have optical counterparts with magnitude $6-15$, of which there are 3699
catalogued\footnote{\url{http://www.hs.uni-hamburg.de/EN/For/Kat/radiost.html}}.
These stars could be first observed in a snapshot survey to measure their radio
brightness and compactness; we expect to find hundreds of radio stars as targets 
for parallax measurements. As a next step, suitable mJy-level calibrators 
will be sought within a few arcminutes of the targets. All this work can be
carried out with 50\% SKA1-MID capabilities in a modest VLBI array 
(Table~\ref{tab:ska-vlbi}). It is expected that in phase~1, SKA-VLBI will provide
parallaxes with $<$10~$\mu$as accuracy for hundreds of stars. To calibrate
the {\it Gaia} parallax zero point with an uncertainty of 0.6 $\mu$as, we will
need parallax measurements for $\sim$500 stars.

\subsection{Possible science applications from comparing radio \&
optical astrometry}

SKA-VLBI will significantly improve the connection between the celestial 
reference frames defined in the optical and radio bands. Currently, the most 
accurate fundamental celestial reference frame is ICRF2, the second realization 
of the International Celestial Reference Frame \citep{Fey2009}, constructed
with dual-frequency (2.3/8.4~GHz) VLBI observations of selected
radio-loud AGN. By about 2020, {\it Gaia} is expected to construct a reference
frame with an accuracy similar to or better than that of VLBI, based directly 
on optical AGN measurements. For maintaining the consistency between the optical
and radio frames, it will be essential to align the {\it Gaia} and VLBI frames
with as many common high-quality reference objects as possible. However,
the number of known optically bright AGN detectable with {\it Gaia}
that also have compact mas-scale radio structures is low~\citep[e.g.][]{Bourda08}. 
Increasing this number depends largely on having deeper VLBI observations 
\citep{Bourda10}, and the high sensitivity SKA-VLBI on long baselines will
play an important role here, at least in the southern sky 
(see Sect.~\ref{sec:surveys}). 

Pulsar -- white dwarf binary systems provide further opportunity for reference 
frame ties with {\it Gaia}. There are currently $\sim$120 binary systems known 
that consist of a white dwarf and a radio pulsar \citep{manchester05}. With the
SKA pulsar surveys that number could increase 5-fold. Comparison of 
the positions derived from {\it Gaia} and SKA VLBI would considerably enhance the 
reference frame tie provided by radio stars alone. Importantly, the majority of
pulsars with a white dwarf companion are millisecond pulsars, for which 
$\sim\mu$as-precision position measurements are possible from timing data alone 
\citep{smits11}.
Pulsar timing positions are based on the dynamic solar system frame, while
interferometric positions are based on the inertial quasar reference frame 
therefore pulsars can be used to tie the different reference frames \citep{bartel96}. 
By combining positions derived from pulsar timing data with positions obtained 
with {\it Gaia} and SKA-VLBI, the pulsar-white dwarf binary systems will tie the 
three reference frames ({\it Gaia} frame, the quasar frame and solar system
dynamical frame) with better than 10~$\mu$as precision.

To align the radio and optical reference frames using AGN as described 
above, one has to take into account that the radio and optical peak-brightness
positions of compact AGN are not necessarily coincident at the accuracy offered 
by {\it Gaia} and VLBI, a phenomenon known as \lq\lq core shift''.
To measure this effect requires observations over a wide range of frequencies, preferably 
extending to even above 10~GHz. With comparable astrometric accuracies in the radio 
and optical, {\it Gaia} will provide important constraints to this by accurately 
locating the position of the central engine with respect to the observed radio structure. 
SKA2 will be able to measure core shifts on sub-mas scales for large sample of AGN 
and thereby constrain AGN jet properties such as the magnetic field strength, 
the non-thermal particle density, jet power \citep{Lobanov98,Kovalev08} and even the 
accretion physics \citep{Zamaninasab14}. The measurement of the jet power, combined with 
data at X-ray energies and/or at optical wavelengths, could provide an estimate of the 
accretion rate and black hole spin \citep{Martinez-Sansigre11}. 

Comparing {\it Gaia} and SKA-VLBI positions of the centres of galaxies will be a great
tool for finding astrophysically interesting objects as well. Off-centre compact radio 
sources could be related to background quasars gravitationally lensed by the galaxy, or 
indicate dual active nuclei in merging galaxies where one of the components is 
radio-quiet while the other is optically obscured but radio-loud \citep[e.g.][]{Orosz13}. 
The latter might be expected for example in minor mergers because the smaller black hole 
will undergo enhanced accretion episodes during the merger process \citep{Callegari11}. 
A particularly interesting case would be the identification of recoiling supermassive 
black holes that are expelled from the centre of the host galaxy due to three-body 
interactions \citep{Hoffman07}, or gravitational-wave recoil after binary coalescence 
in a galaxy merger \citep{Blecha11,Komossa12}. 

\begin{figure} 
\centering
\includegraphics[scale=0.2, viewport=14 14 1634 1727, clip]{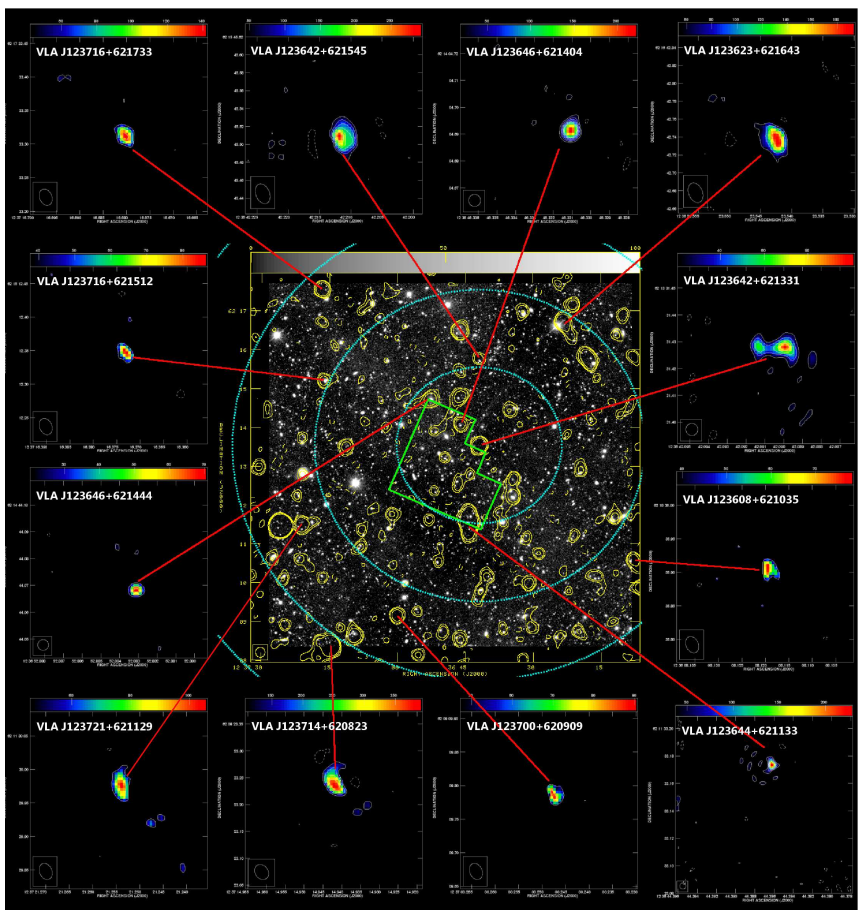}
\caption{VLBI-detected AGN in the field of 
Hubble Deep Field North and Flanking Fields
(grey scale: optical; yellow contours: WSRT 1.4 GHz). The cyan circles represent 
annuli of decreasing resolution and sensitivity, and are drawn at 2, 4, 6, and 8 
arcmin radius w.r.t. the phase center which coincides with radio AGN 
VLA J123642+621331 \citep[][Fig.~1]{Chi13}.    
\label{fig:HDF-N}}
\end{figure}

\section{SKA-VLBI surveys} \label{sec:surveys}

Large field of view (FoV) VLBI surveys have a number of important applications like  
exploring the AGN content of the universe \citep[identified by their observed high 
brightness temperature, e.g.][see Fig.~\ref{fig:HDF-N}]{Chi13} and revealing 
the AGN physical properties down to very low accretion rates 
($10^{-6}-10^{-9} L_{\rm Edd}$) and masses ($<10^6 M_{\odot}$). 
Most importantly, combining
radio data products with the the \lq\lq multicolor'' sky view of the Large Synoptic 
Survey Telescope (LSST) will provide invaluable information on a huge number of 
individual objects; for example, accurate photometric redshifts will be available 
for millions of faint targets out to $z>2$, essential to understanding the evolution
of activity in the universe as traced by radio emission \citep{Lazio14}. Although at
present, the FoV in VLBI observations has been strongly limited by 
data averaging in time and frequency, new techniques promise to allow the VLBI 
imaging of sources over most or all of the primary beam.
Wide-field observations to map the full field of view
of the primary beam of individual telescopes have been carried out, but this
was a computationally challenging exercise \citep[e.g.][]{Lenc08}.  The use of
fast hierarchical widefield mapping procedures can reduce the computational
burden significantly \citep{Wucknitz10}.  However, in recent years the advent
of ``multifield" correlation in VLBI software correlators like DiFX \citep{AdamDifx} 
have provided an alternative and efficient means of imaging multiple targets
within the FoV of the VLBI array. Internal to the correlator, visibilities are
processed with high frequency resolution, and are shifted to the positions
of the target sources with a high cadence ($\sim$100 Hz) before being
heavily averaged \citep{Morgan11}. In this way
hundreds of previously-identified sources can be studied in a single 
observation at mas resolution, without the generation of an extremely large dataset.

The main limitation to SKA-VLBI survey science will be the small number 
of tied-array beams (N$\sim$4) available with SKA1. The relatively 
sparse nature of the SKA1-MID and SKA1-SUR cores means that only
a small fraction of the full primary field of view will be visible to the 
tied array beams.  A larger tied-array beam can be formed at the expense
of sensitivity, by including fewer antennas to a shorter maximum diameter.
In most cases, however, optimal sensitivity will be gained by using sensitive, narrow
tied-array beams and switching between targets with a rapid cadence.  Standard
multi-field VLBI correlation targeting all sources within the primary beam would
be used (since all targets would be visible at all times to the other, single-dish elements
of the SKA-VLBI array) but only a subset of the targets would have a tied SKA
beam contributing sensitivity at any given time.
An important issue is how to calibrate the gain of the remote antennas accurately in the 
direction of the chosen VLBI phase centres, since the primary beam response 
of the individual telescopes is often poorly known \citep[][]{Middelberg11,Cao14,Deller14}. 
This is unlikely to be a concern for tied array beams from SKA1-MID and SKA1-SUR, 
however, since the primary FoV of the dishes will be order of $1^{\circ}$, considerably 
larger than some other elements of the SKA-VLBI array (and hence the area 
from which targets will be drawn).

SKA2 will be capable of streaming and processing data from all
individual elements to provide high resolution data for practically the full 
FoV, although it may still only be practical to provide the full resolution for
limited areas, e.g. hundreds of independent phase-centers. The full
SKA will thus provide a unique range of mas to arcsec angular scales that will 
help to distinguish between various interpretations of faint radio sources 
(unresolved AGN core, resolved emission powered by an AGN, star forming 
complexes etc.).

\section{Resolving explosive outflows} \label{sec:GRBjets}

SKA-VLBI will make possible (sub)-mas imaging of radio sources. In particular, 
for transient phenomena such us explosive outflows, SKA-VLBI monitoring will allow 
us to measure the expansion velocity, or proper motion of jetted ejecta, which is 
important to understanding the physics of the explosion as well as for studying 
the progenitor environments. The various Galactic (e.g. microquasars, magnetars, 
novae) and extragalactic (e.g. TDE)  
examples are discussed in other chapters. Here we describe a particular application 
of SKA-VLBI, as an example, in order to highlight the importance of improved 
sensitivity in global VLBI arrays.

Some of the most energetic events in the universe, supernovae (SNe)
and long-duration gamma-ray bursts, involve explosive outflows of material
from dying stars (the latter observed in the other parts of the electromagnetic 
spectrum as GRB afterglows). Supernovae are not only intrinsically spectacular,
but they also have a significant effect on both the chemical evolution
of galaxies and inject significant amounts of energy into the
interstellar medium.  Some supernovae, specifically those of Type Ib and
Ic, which originate in stars that have lost most of their original
H-rich stellar envelopes, can eject material at relativistic speeds.
Long duration GRBs are known to be associated with 
Type Ic supernovae, and thought to be produced by ultra-relativistic jets
launched in the collapsing star.
Many details about these processes are still unknown.  Why do only a
small fraction of Type~Ic SNe give rise to an observed GRB?  How is the
relativistic jet launched and how does it evolve?  VLBI provides
the best way of obtaining spatially resolved information about
these sources, which would be of obvious benefit in attempting to understand
the physics.  Radio observations have the further advantage that radio emission 
generally traces the fastest outflows in GRB afterglows.

VLBI observations of GRBs and SNe are limited both by the available
resolution and sensitivity.  
Although the SKA will provide no direct increase
in resolution, it will provide an indirect increase in that the 
accuracy with which model-fitting can be used to constrain
source sizes increases with increasing signal-to-noise ratio (SNR).
In SKA1-MID Band 4--5, with a global array of VLBI telescopes where the 
longest baselines are $\sim$10,000~km (approximately the distance between the
SKA core in South Africa and Australia or Europe), native imaging
resolutions down to $\sim$1~mas (FWHM) would be obtained.
For SNR$\sim$50, the minimum detectable angular diameter is 
$\sim$0.2~mas\footnote{  
The theoretical resolving power of an interferometer is inversely proportional 
to the signal-to-noise ratio. The minimum detectable size depends weakly on the 
source geometry, and can vary by $\sim$30\% for different source geometries such 
as a uniform disk or an optically thin sphere \citep{Marti-VidalPL2012}.}.
This minimum detectable size corresponds to a length of $\sim$11
light-days ($3\times10^{16}$ cm) at distance 10 Mpc, and $\sim$1.6
light-years ($8\times10^{18}$ cm) at a redshift, $z$ of 0.15.
Interestingly, at $z = 20$, it corresponds to the same length of $\sim$1.6
light years, so sources at very large cosmological distances can be
resolved provided they are bright enough to be detected.
Such measurements therefore have the exciting potential to directly
measure the expansion of a nearby relativistic outflow in the first
two weeks, and to possibly resolve outflows even at cosmological
distances. 

\begin{figure}
\includegraphics[scale=0.35]{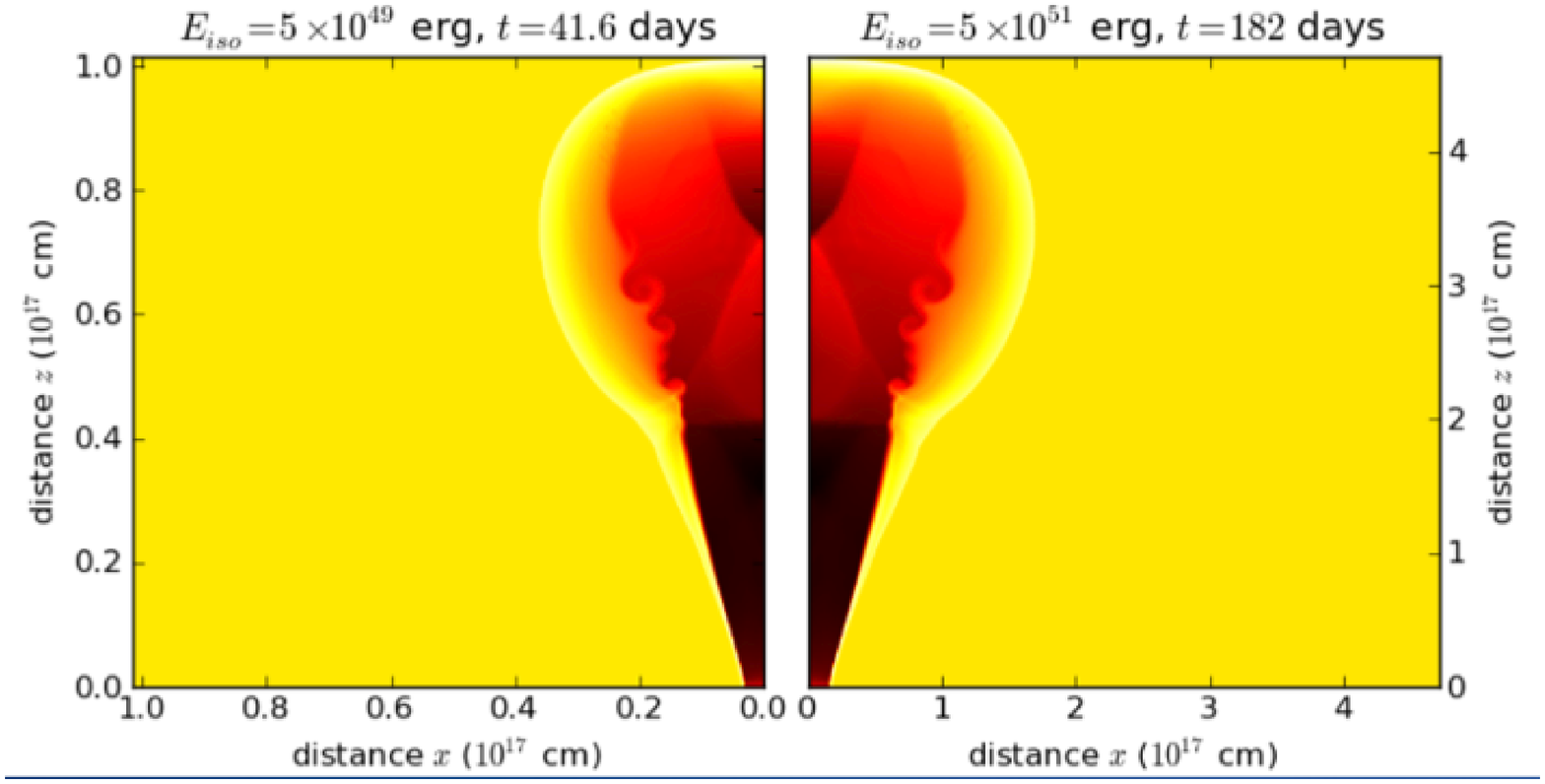}
\includegraphics[scale=0.25, viewport = 0 0 535 510, clip]{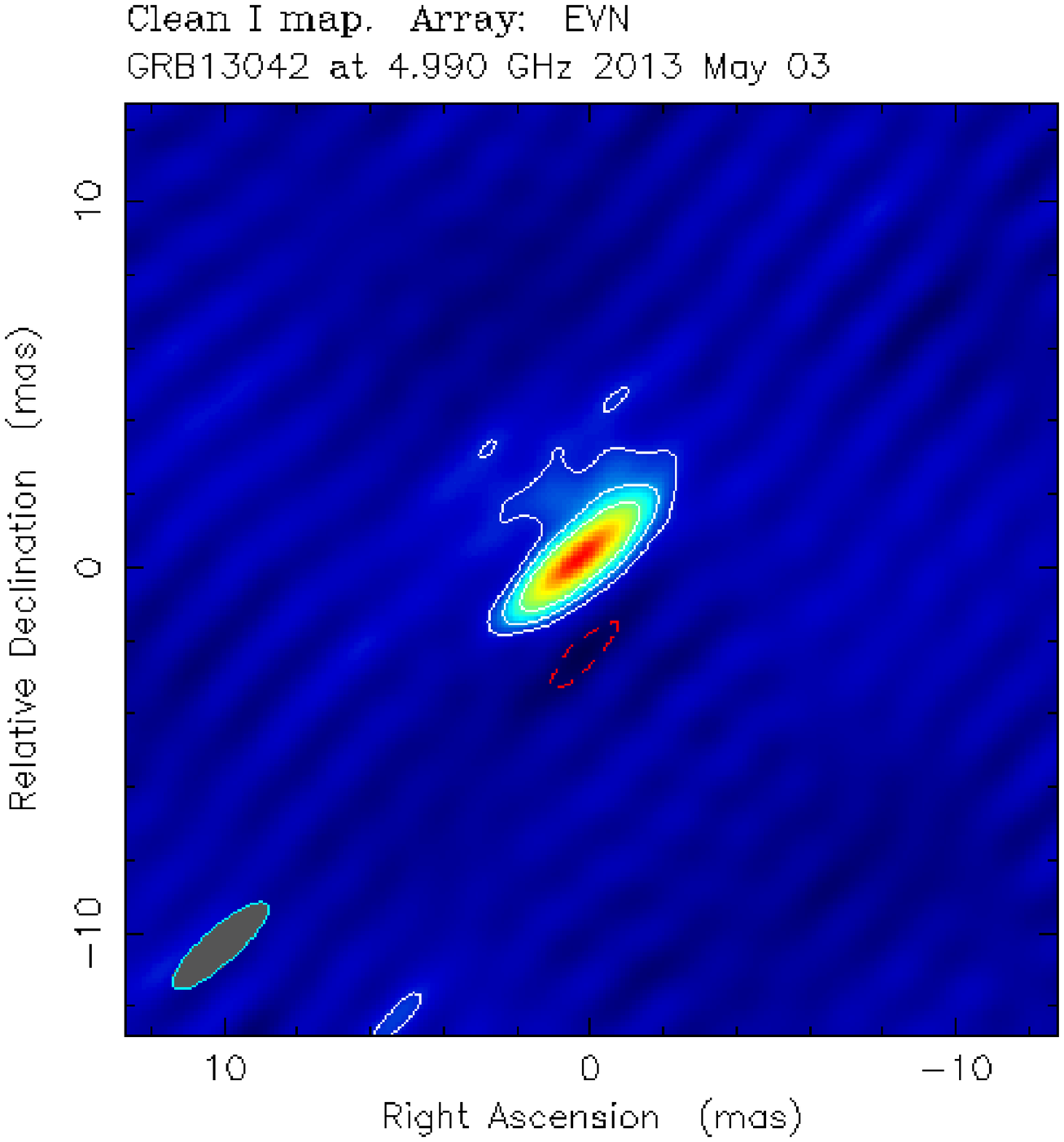}
\caption{
\small Left: 2D simulations of gamma-ray burst (GRB) jets. The scale invariance
between relativistic jets of different energy allows easy reproduction of the jet dynamics.
This provides a powerful method for fitting simulation results to observational data 
\citep{vanEerten12}. With SKA-VLBI this will be possible for a large number of GRBs. 
Figure courtesy of Hendrik van Eerten. Right: Early e-EVN detection of the highest fluence 
GRB detected for decades, GRB~130427A \citep{Paragi13}.
\label{fig:GRB}}
\end{figure}

SKA-VLBI will allow us to observe sources 
below the detection limits of current instruments. Of all GRB afterglows,
only $\sim$30\% are bright enough to have been detected in the radio
with present instrumentation \citep{ChandraF2012}, and only GRB 030329
has been resolved \citep[e.g.][]{Taylor+2004}. Similarly, only $\sim$30\% of 
the observed core-collapse supernovae were detected in the radio, and only 
a handful have well-resolved images \citep{Bietenholz2008, Bietenholz2014}.  
The sensitivity of SKA would allow a significantly increased sample of 
both GRBs and SNe to be resolved with VLBI. We note that 
the minimum detectable source
size could be significantly smaller than 0.1~mas for SNR$\sim$1000. This SNR
should be achieved for the most extreme and nearby $>$1~mJy sources with SKA1, 
and with SKA2 for practically all GRB afterglows with a flux density of 
a few hundred $\mu$Jy known as \lq\lq radio-loud'' today (cf. Table~\ref{tab:ska-vlbi}). 
The resulting large GRB samples will give the first opportunity to study 
GRBs in a model-independent fashion while they are still in their ultra-relativistic 
phase (that can last up to about two weeks for sources that are in a low-density 
environment), and directly compare the results with detailed simulations (see 
Fig.~\ref{fig:GRB}). This will however require extremely accurate calibration for 
all interferometer elements and flexible response to triggers;
detailed simulation of SKA-VLBI configurations will provide more robust estimate 
of realistic expectations for \lq\lq super-resolution'' capability with SKA-VLBI.

A particularly exciting possibility is that of detecting and resolving
the outflows from GRBs from the first generation of stars ---
Population III GRBs.  GRB afterglows have already been detected at $z > 9$
\citep{Cucchiara+2011}, and recent work \citep[e.g.][]{Ghirlanda+2014,
  Mesler+2014} has shown that the very massive stars expected to form
in the early universe could give rise to spectacular GRBs, and their afterglows 
may be detectable by SKA despite being at $z \sim 20$.

\section{SKA in VLBI observations: data transport, data formats and correlation} \label{sec:data} 

\subsection{Status of global VLBI}
The EVN currently operates at a bandwidth of 128~MHz and 2-bit sampling, resulting in data 
streams of 1~Gbit~s$^{-1}$ per telescope. The VLBA in the USA has recently completed an upgrade to 
2~Gbit~s$^{-1}$. The ongoing roll-out of new digital backends (DBBC) in the EVN will make 
2 and 4~Gbit~s$^{-1}$ operations possible. After the completion of this 
roll-out, the only obstacles for high-bandwidth global VLBI will be the availability of 
suitable receiver systems (feeds etc.) at the telescopes, and of sufficient magnetic media 
for recording or sufficient networking bandwidth for real-time correlation of the data streams.
The RFI environment at the different telescope sites will of course limit the usable bandwidth.
Using phased-up arrays of radio telescopes (like the ATCA, the VLA and the WSRT) as elements 
in VLBI observations is common practice. Once beam forming has taken place, the output of the
array is treated like that of any of the other VLBI stations.  

\subsection{SKA-VLBI}
The cores of SKA1-MID and SKA1-SUR will produce very narrow beams. The practice of using
in-beam calibrators in order to obtain the highest quality phase calibration will only be 
possible if at least four beams are formed, one for the target and additional three for the
calibrators (see Sect.~\ref{sec:calibration}). With the usual caveats, one can 
reasonably assume that on the timescale
of the construction of SKA1 most VLBI telescopes will move to 512~MHz bandwidth 
operations. Assuming 4 beams, 512 MHz bandwidth and 2-bit sampling, the total data rate of 
the phased-up SKA would be 16~Gbit~s$^{-1}$. While current software correlators can easily deal with 
different bit representations, and a higher bit representation of the SKA data would yield 
some additional gain in sensitivity, it probably would make most sense to re-sample/truncate 
(depending on the exact representation) the SKA data at the telescope to 2~bits/sample, thus 
saving on data volume and storage/transport. The SKA data would also have to conform to the 
standard VLBI subband scheme, meaning that subbands of 32~MHz should be available. Care should 
be taken that the band can be tuned so as to be compatible with standard VLBI settings. 

\subsection{Data format}

VLBI has known a variety of data formats. In recent years however a common format named VDIF 
\citep{whitney09} has been accepted by most of the community. Most newly developed VLBI equipment
now supports this data format. While dealing with different formats nowadays is far less 
problematic than in the days of ASIC-based hardware correlators, one common format is of course 
preferable. It is not yet known what SKA data will look like, but after beam forming, truncating 
and maybe re-filtering it should be possible to pack the data in any format desired. Related to 
this, the SPEAD \citep{manley07} protocol which will be used for data exchange in MeerKAT, and
maybe for SKA1-MID as well, actually supports the VDIF format. Hartebeesthoek Observatory in 
South Africa has participated in a successful 4~Gbit~s$^{-1}$ e-VLBI demonstration, streaming data in 
real time to the EVN correlator at JIVE in the Netherlands. Considering that 100~Gbit~s$^{-1}$ networking 
technology is being rolled out by national and international research networks, transporting a 
data stream of 16~Gbit~s$^{-1}$ from South Africa or Australia to Europe (or another 
correlator location) in real-time should be no problem at all in five years. Recording such data 
streams and trickling them to a correlator at lower speeds at a later time (\lq\lq e-shipping'') 
is already possible right now.

Finally, many VLBI arrays are configured using the VEX 
(VLBI EXperiment)\footnote{http://www.vlbi.org/vex/docs/vex$\%20$definition$\%20$15b1.pdf} 
format, providing a complete description of a VLBI experiment, including scheduling, 
data-taking and correlation. Some sort of interface will be needed to translate VEX 
files into configuration files understandable by the SKA control system.

\section{Conclusions}

In this paper we described the scientific motivation and possible technical realisations 
of SKA-VLBI. The science goals are best achieved with SKA1 by forming phased-array 
elements from SKA1-MID and SKA1-SUR observing together with existing VLBI arrays in the 
1-15 GHz frequency range (and up to 22~GHz in SKA2).
In addition, the combination of local interferometry data from SKA1-MID and SKA-VLBI
will provide the basis for very accurate amplitude and polarization calibration of 
the VLBI data products. A high resolution configuration of SKA2 (with a resolution 
of tens of mas to $\sim$100 mas, depending on frequency) will allow imaging of e.g. 
extragalactic sources from sub-pc to kpc scales simultaneously, a capability that is
rarely available today. An important part of the requirements for the SKA is forming
multiple VLBI beams for accurate phase calibration, targeting multiple sources within
the primary beam. This mode of operation will be compatible with other multi-beam
VLBI components such as the WSRT-APERTIF. 

VLBI with the SKA1 will not be very different from that with other 
tied arrays, in terms of operations, data rates and computing. More beams will mean more data, but 
the numbers are reasonable, even in relation to today's technology. As observing at 
higher frequencies becomes possible with the SKA, increasing the instantaneous bandwidth 
will become increasingly important. Higher bandwidths, and the consequent higher data
rates, should become feasible in the relevant time-frame. 

SKA-VLBI will have a profound effect on a large number of fields within astronomy.
Accreting objects will be in reach at a range of accretion rates well below Eddington,
providing a comprehensive view of compact objects such as neutron stars as well as 
the full range of masses from stellar- to supermassive black holes. Many of 
these will be discovered in transient surveys. Especially interesting candidates are
TDEs that could provide clues to derive the low end of the SMBH mass-function, and 
extremely high redshift GRBs that would provide a line of sight through a large volume 
of the universe.
These, along with the great number of newly discovered dual and multiple SMBH systems 
and recoiling BH will provide clues for understanding structure formation in the 
early universe. In addition, there will be a wealth of information about the feedback 
processes between the central AGN and their host galaxies from spectral line VLBI 
surveys at $\sim$1~GHz. At higher frequencies the study of star formation and 
stellar evolution, as well as resolving explosive outflows in the Galaxy and at 
cosmological distances will be among the most important applications. As explained 
in this work, VLBI astrometry will remain a very important tool for astrophysics. 
For example pulsar parallax measurements using SKA-VLBI will play an essential role 
in several high impact areas, including strong field tests of gravity in relativistic 
binary systems, tomographic mapping of the Galactic magnetic field and mapping the 
ionised interstellar plasma in the Galaxy, and the physics of 
neutron stars, as well as detecting the gravitational wave background. A particularly 
interesting idea is the detection of the gravitational wave background directly through 
measuring proper motions of a million of quasars with SKA2.

\setlength{\bibsep}{0.0pt}
\bibliographystyle{apj}

\newcommand{\amp}{\&}
\newcommand{\araa}{ARA\&A}
\newcommand{\aap}{A\&A}
\newcommand{\aapr}{Astron. Astrophys. Rev.}
\newcommand{\aaps}{A\&A Suppl.}
\newcommand{\adspr}{Adv. Sp. Res.}
\newcommand{\aj}{AJ}
\newcommand{\apj}{ApJ}
\newcommand{\apjl}{ApJL}
\newcommand{\apjs}{ApJS}
\newcommand{\apss}{ApSS}
\newcommand{\baas}{BAAS}
\newcommand{\memras}{Mem. R. Astron. Soc.}
\newcommand{\memsai}{Mem. Soc. Astron. Ital.}
\newcommand{\mnras}{MNRAS}
\newcommand{\iaucirc}{IAU Circ.}
\newcommand{\ijmpd}{Int. J. Mod. Phys. D}
\newcommand{\jrasc}{J. R. Astron. Soc. Can.}
\newcommand\nar{New Astronomy Reviews}
\newcommand{\nat}{Nat}
\newcommand{\pasa}{PASA}
\newcommand{\pasj}{PASJ}
\newcommand{\pasp}{PASP}
\newcommand{\sci}{Sci}

\end{document}